\documentclass[twocolumn]{revtex4}
\usepackage[dvips]{graphicx}
\usepackage{float}

\begin{document}

\title{Model for magnetic flux pattern induced by in-plane magnetic fields on spatially
ÊʎÊinhomogeneous superconducting interfaces of LaAlO$_3$-SrTiO$_3$}

\author{Kazushi Aoyama ${}^{1,2,3}$ and Manfred Sigrist ${}^{2}$ }

\affiliation{${}^1$ The Hakubi Center for Advanced Research, Kyoto University, Kyoto 606-8501, Japan \\
${}^2$ Institute for Theoretical Physics, ETH Zurich, Zurich 8093, Switzerland \\
${}^3$ Department of Physics, Kyoto University, Kyoto 606-8502, Japan }

\begin{abstract}
The effect of spatial inhomogeneity on the properties of a two-dimensional non-centrosymmetric superconductor in an in-plane magnetic field is investigated, as it can be realized in LaAlO${}_3$-SrTiO${}_3$ interfaces. 
We demonstrate that the spatial variation of Rashba spin-orbit coupling (RSOC) yields a local magnetic flux pattern due to the field-induced inhomogeneous helical phase. For sufficiently strong fields, 
vortices can nucleate at inhomogeneities of the RSOC. 
\end{abstract}

\maketitle

The magneto-electric effects belong to the most intriguing features of superconductors without inversion symmetry \cite{normal_jM, SC_jM, SC_Bj, Dimitrova,Samokhin,Kaur,Fujimoto}. The connection between spin polarization and supercurrents is one basic aspect which has numerous consequences. For example, spin polarization, by external magnetic fields or intrinsic ferromagnetic order, can induce a so-called helical phase in the superconducting state. Such a helical phase can appear as simple phase factor 
$ e^{\i {\bf q} \cdot {\bf r}} $ to the superconducting order parameter, resembling a Fulde-Ferrell-Larkin-Ovchinnikov (FFLO) \cite{FF, LO} state
\cite{Dimitrova,Samokhin,Kaur,Springer,Agterberg}. However, also amplitude modulations of the order parameters are possible on this basis \cite{Kaur,Ikeda_0, Ikeda}. Such magneto-electric effects originate from spin-orbit coupling specific to non-centrosymmetric metals. 

Non-centrosymmetric superconductors (NCSC) most extensively investigated are those lacking mirror symmetry with respect to a crystalline plane (e.g.: $a$-$b$-plane of a tetragonal crystal) and possess spin-orbit coupling of the Rashba-type with the well-known structure: $ \alpha ({\bf k} \times {\hat z}) \cdot {\bf S} = \alpha {\bf g}_{\bf k} \cdot {\bf S} $ \cite{Rashba}. Non-centrosymmetricity of this kind is realized in some heavy Fermion superconductors such as CePt$_3$Si \cite{CePt3Si} and CeTSi$_3$ (T=Rh, Ir) \cite{CeRhSi3, CeIrSi3} and in superconducting thin films grown on a substrate. A particularly interesting example of the latter case is
the superconductor found at interfaces between the band insulators LaAlO$_3$ and SrTiO$_3$ \cite{interface-SC}. Here, superconductivity and spin-orbit coupling can be influenced by perpendicular electrical gate fields \cite{tunable-SC, tunable-RSO}.
This system provides a unique platform to study magneto-electric effects. Recently, scanning SQUID experiments have detected local magnetic patterns which have been interpreted as ferromagnetic patches coexisting with the superconducting phase \cite{imaging}. 
Motivated by these experiments and the desire to detect signatures of the so far elusive helical SC phase \cite{Kaur}, we investigate magnetic properties of a 2D superconductor in an inhomogeneous environment exposed to an in-plane magnetic field or magnetization.

\begin{figure}[t]
\begin{center}
\includegraphics[scale=0.45]{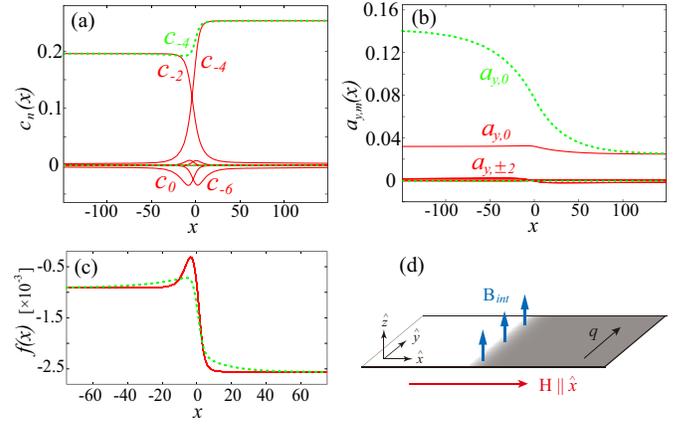}
\caption{(Color online) Spatial dependence of the Fourier components of the SC gap function $c_{n}(x)$ (a) and the vector potential $a_{y,m}(x)$ (b) along the $x$-axis in an inhomogeneous state [solid (red) curves] with and a homogeneous state [dotted (green) ones] without spatial modulation in the $y$ direction at $T/T_c=0.9$, $H/H_{p,-}=0.976$. (c) The free-energy densities of the two states are averaged over the $y$-range. The length are given in units of $\xi(t)$ with $d/\xi(t)=3.0$ and $\lambda_0(T)/\xi(T)=9.0$. (d) Geometry of the inhomogeneous 2D NCSC with strong (weak) RSOC on the right (left) half-plane
whereby an in-plane magnetic field ${\bf H}|| {\hat x}$ yields out-of-plane magnetic flux ${\bf B}_{int}||{\hat z}$ around the boundary (see Fig. \ref{fig:flux}).\label{fig:x-dep}}
\end{center}
\end{figure}

A field parallel to the plane of a sufficiently thin film of a NCSC would generate an ideal helical phase without being disturbed by vortex lattice modulations. However, this phase would not leave obvious traces as its phase gradient can be compensated by a gauge field equivalent to the screening of magnetic fields in superconductors \cite{Springer}. The situation changes, if the NCSC is inhomogeneous in this plane, for instance, through modulation of spin-orbit coupling. We will show below on a simple geometry of two half-planes of different spin-orbit coupling strength, how in the helical phase magnetic flux pattern and even vortices oriented perpendicularly to the film can appear in the region of varying spin-orbit coupling (see Fig. \ref{fig:x-dep} (d)). 


We model our system by the BCS-like Hamiltonian,
\begin{eqnarray}\label{eq:Hamiltonian}
&& {\cal H} = \sum_{{\bf k},s,s'} K_{ss'} ({\bf k}) c_{{\bf k}s}^{\dag} c_{{\bf k},s'} - a  \sum_{\bf q} \, B^\dagger({\bf q}) B({\bf q}),\nonumber\\
&& K_{ss'}({\bf k}) = \varepsilon_{\bf k} \delta_{ss'}+ {\mbox {\boldmath $\sigma$}}_{ss'}\cdot \big[ \alpha \, {\bf g}_{\bf k} + g\mu_B {\bf H}\big],
\end{eqnarray}
where $c_{{\bf k} s}$ annihilates a quasiparticle state with  momentum ${\bf k}$ and spin $s$, 
and 
\begin{equation}
B({\bf q}) = \frac{1}{2} \sum_{{\bf k},s,s'} (-i \, \sigma_2)_{ss'} \, c_{-{\bf k}+\frac{\bf q}{2}s} c_{{\bf k}+\frac{\bf q}{2}s'} .
\end{equation}
The single-electron spectrum is described by $ K_{ss'} $ including the RSOC with strength $ \alpha  $ and the unit vector $ {\bf g}_{\bf k} = \hat{\bf k} \times \hat{z} $ ($ \hat{\bf k} = {\bf k}/|{\bf k}| $) and
the Zeeman field $  g\mu_B {\bf H} $ coupling to the spin ($ g $: $g$-factor and $ \mu_B $: Bohr magneton). The kinetic energy $ \varepsilon_{\bf k} $ is measured from the Fermi energy $ \varepsilon_F $. Finally, we restrict ourselves to a pairing interaction of strength $a$ ($>0 $) in the (onsite) s-wave channel\cite{pairing}.

The free energy derived from the above model can be expanded in powers of the gap function $ \Delta $ 
($ \Delta({\bf q}) = a \langle B({\bf q}) \rangle $) to yield a generalized Ginzburg-Landau (GL) theory valid near $ T_c $ \cite{Dimitrova, Samokhin, Kaur,Springer}.
\begin{eqnarray} 
{\cal F}_{GL} & = & \int d^2r \left[ A_2 | \Delta |^2 + A_4  | \Delta |^4 + K |  {\bf \Pi} \Delta|^2  \right. \\
    && \left. + \tilde{K} (\hat{z} \times {\bf H} ) \cdot \{ \Delta^* {\bf \Pi} \Delta + \Delta ({\bf \Pi} \Delta)^* \} + \frac{ (\nabla \times {\bf A})^2}{8 \pi} \right] , \nonumber
 \label{eq:GL}
 \end{eqnarray}   
where the coefficients are given by
\begin{equation}
A_2 = N_0 \left( \ln \frac{T}{T_c} + 2 \gamma g^2 \mu_B^2 H^2 \right), 
\end{equation}
$ A_4 = N_0 \gamma $, $  K= N_0 \gamma v_F^2 /2 $, and $ \tilde{K} = \delta N_0 g \mu_B v_F \gamma /2$
with $ \gamma = 7 \zeta(3)/16 (\pi k_B T_c)^2 $, $ N_0 = (N_+ + N_-)/2 $ as the mean density of states per spin,
and $ \delta N_0 = N_+ - N_- = 2 \alpha N'_0 $ ($ N_{\pm} $; density of states of the two bands split by the RSOC: $ \varepsilon_{\bf k}^{(\pm)} = \varepsilon_{\bf k} \pm | \alpha {\bf g}_{\bf k} | $ and $ N'_0 $ a measure for the particle-hole asymmetry of the Fermi surface). The covariant gradient is given as $ {\bf \Pi} = - i \hbar \nabla + (2e/c) {\bf A} $, where $ {\bf A}$ is the vector potential with $\nabla \times {\bf A}= {\bf B}_{int}$ and ${\bf B}_{int}$ as the internal magnetic field in contrast to the Zeeman field $ {\bf H} $. 
The second term in $A_2$ describes the paramagnetic limiting effect through the Zeeman field $ {\bf H} $. The second gradient term involves 
magneto-electric effects arising from in-plane spin polarization. 
We neglect any spatial dependence perpendicular to the plane assuming the thickness of the SC interface in LaAlO${}_3$-SrTiO${}_3$  sufficiently small \cite{interface-SC, interface-field}. Moreover, we restrict to $|\alpha|/E_F \ll 1$, and spatial variations of $ \alpha({\bf r}) $ on scales larger than the Fermi wave length are incorporated by a straightforward generalization into the GL free energy. 

Before discussing the effects of these terms on the magnetic properties of an inhomogeneous superconductor, we would like to briefly mention one important aspect of the magneto-electric effect and the corresponding helical phase. In a thin film with an in-plane magnetic field $ {\bf H} $
paramagnetic limiting is the relevant pair-breaking mechanism, as orbital pair-breaking is suppressed. 
The renormalized upper critical field is given by
\begin{equation}\label{eq:Hc2}
H_{p}(T) = \frac{H_{p,0} (T) }{\sqrt{1 - \delta N_0^2/(4N_0^2)}}
\end{equation}
with a gap function $ \Delta ({\bf r}) = \Delta_0 \, {\rm exp}(i {\bf q} \cdot {\bf r}) $, the helical vector defined as $ {\bf q}  = - (\tilde{K}/K \hbar) (\hat{z} \times H) - (2e/\hbar c) {\bf A} $, and the bare paramagnetic limiting field
$ H_{p,0} = \sqrt{|\ln (T/T_c)|}/ (g \mu_B \sqrt{2\gamma} )$. Obviously, the wave vector $ {\bf q} $ is gauge dependent, whereas resulting measurable quantities, such as the current densities $ {\bf j}_s + {\bf j}_L$ defined below, are gauge invariant. 
The magneto-electric coupling stabilizes the superconducting state and
increases the paramagnetic limiting field \cite{Kaur,Springer}. Consequently, the inhomogeneity of RSOC would yield inhomogeneous nucleation of superconductivity in a uniform field even in systems which would show
homogeneous superconductivity in zero field.

A first idea of the magneto-electric effect can be gained by considering the GL equation obtained by variation of $ {\cal F}_{GL} $ with respect to $ {\bf A} $ 
\begin{equation}
\nabla \times {\bf B}_{int} = \frac{4 \pi}{c} ({\bf j}_s + {\bf j}_L + {\bf j}_{me} )
\label{pre-London}
\end{equation}
with current densities $ {\bf j}_s = i 2eK \hbar \{ \Delta^* \nabla \Delta - c.c.\} $ and $ {\bf j}_L = - (8e^2 K/c) | \Delta|^2 {\bf A} $, and the magneto-electric contribution $ {\bf j}_{me} = - 2e \tilde{K} | \Delta|^2 (\hat{z} \times {\bf H})$.
While $ {\bf j}_s $ and $ {\bf j}_L $ are standard currents due to phase gradient and London screening contributions, respectively, the last one is characteristic for magneto-electric phenomena due to the field-induced shifts of the Fermi surfaces \cite{Kaur,Springer}. Note that for the uniform helical phase, the sum $ {\bf j}_s + {\bf j}_{me} + {\bf j}_{L}$ vanishes. 
The curl of Eq.(\ref{pre-London}) leads to the London equation
\begin{equation}
\nabla^2 {\bf B}_{int} - \nabla \times \frac{1}{\lambda^{2}} {\bf A} = - {\bf B}_{eff}
\end{equation}
with
\begin{equation}
{\bf B}_{eff} = \frac{4 \pi}{c} \nabla \times {\bf j}_s - \frac{4 \pi e}{c} \hat{z} \nabla \cdot {\bf H} \tilde{K} | \Delta |^2
\label{eq:b-source}
\end{equation}
and $ \lambda^{-2} = 32 \pi K | \Delta |^2 e^2/c^2 $  as the London penetration depth. We find an effective source field whose first term originates from the ordinary phase gradient, while the second part is anomalous and depends on the spatial dependence of the Zeeman field, the spin-orbit coupling and/or the order parameter size \cite{Neupert}.  

We now turn to the example of inhomogeneous spin-orbit coupling with two half-planes of different magnitude of $ \delta N_0 $ described by
\begin{equation}
\label{eq:dN}
\delta N_0({\bf r})= \delta \bar{N_0} \Big[ 1+  \eta \big(\tanh(x/d)+1\big)\Big],
\end{equation}
with a boundary of thickness $ d $. This gives rise to a difference between $ \tilde{K} $ on the two sides: $ \Delta \tilde{K} = g \mu_B v_F \gamma \eta \delta \bar{N}_0 $. The magnetic field $ {\bf H} =(H,0,0) $ is uniform and directed along the $x$-axis (perpendicular to the domain boundary). We obtain the spatial structures of the SC gap function $\Delta$ and the vector potential ${\bf A}$, by numerically solving the GL equations $\delta {\cal F}_{\rm GL}/\delta \Delta=0$ and $\delta {\cal F}_{\rm GL}/\delta {\bf A}=0$ under suitable boundary conditions at $x=+\infty$. Here, the Fourier transformations are defined by 
$\Delta({\bf r})  =  k_B T_c \sum_n \, c_{n}(x) \exp [i\, 2\pi n y/L_y]$ and 
$ {\bf A} ({\bf r}) =  {\hat y} \, (\hbar c/2e) \xi(T)^{-1} \sum_m a_{y,m}(x) \exp [i\, 2\pi m y/L_y ] $,
and the boundary condition is expressed as
\begin{eqnarray}\label{eq:boundary}
&& c_{n}(+\infty)   = \delta_{n,n_b} \, \frac{g\mu_B H_{p,0}}{k_BT_c}\Big( 1 - \frac{H^2}{H^2_{p,+}}\Big), \nonumber\\
&& a_{y,m}(+\infty) = - \delta_{m,0} \Big(\frac{\delta N_0(+\infty)\, H}{2N_0 \, H_{p,0}} + 2 \pi n_b \frac{\xi(T)}{L_y} \Big) ,
\end{eqnarray}
with $H_{p,\pm}(T) = H_{p,0} (T) /(1 - [\delta N_0(\pm \infty)/2N_0]^2)^{\frac{1}{2}}$, $\xi (T)= \hbar v_F \sqrt{\gamma/2 |\ln(T/T_c)|}$ as the coherence length, and $n_b$ is defined as an integer minimizing $|a_{y,0}(+\infty)|$. 
Here, the parameters $L_y/\xi(T)=115$, $\delta \bar{N}_0/N_0=0.16$, and $\eta=0.75$ are used, and then, $n_b=-4$ at $T/T_c=0.9$, $H/H_{p,-}=0.976$ ($L_y $: extension of the system along the $y$-direction). 

We find two basic phases for the given geometry. One is a state homogeneous along $y$-direction with a single helical vector $q_y (+\infty)$ and the other is an inhomogeneous state consisting of two main helical vectors $q_y(-\infty)$ and $q_y(+\infty)$. Figure \ref{fig:x-dep} shows the numerically obtained spatial profiles of the Fourier components of $\Delta$ (a) and $A_y$ (b) along the $x$-axis in the homogeneous state [dotted (green) curves] and the inhomogeneous one [solid (red) ones]. 

\begin{figure}[t]
\begin{center}
\includegraphics[scale=0.8]{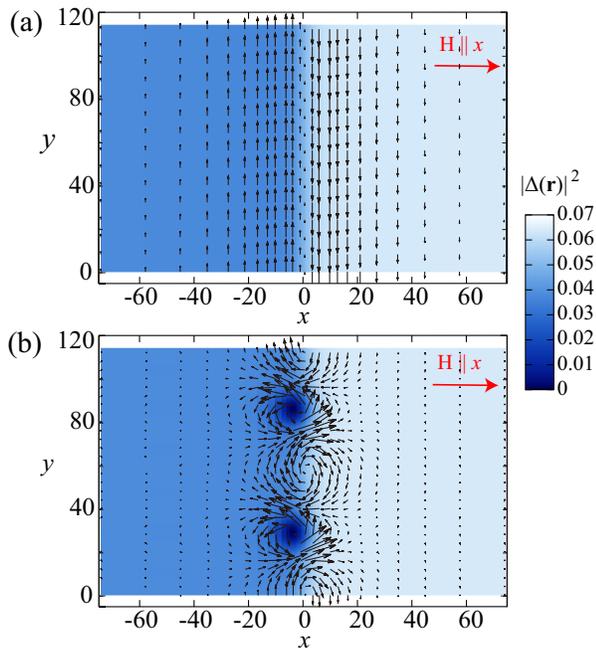}
\caption{(Color online) Spatial structures of the SC gap function $|\Delta|^2$ and the local current ${\bf j}={\bf j}_s+{\bf j}_L+{\bf j}_{em}$ (arrows) in the homogeneous helical state (a) and the inhomogeneous one with a spatial modulation in the $y$ direction (b) (same parameters as in Fig.\ref{fig:x-dep}). The length are given in units of $\xi(t)$.
In (b), vortices threading the interface appear on the boundary at $x=0$ for the in-plane field ${\bf H}||{\hat x}$. \label{fig:gap-current}}
\end{center}
\end{figure}

\begin{figure}[t]
\begin{center}
\includegraphics[scale=0.6]{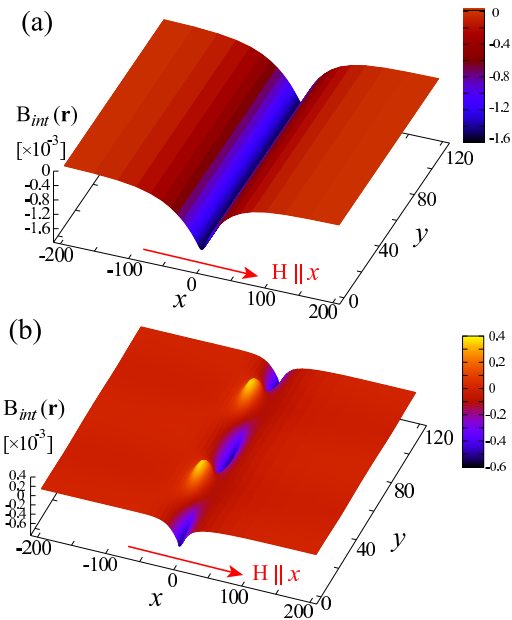}
\caption{(Color online) Spatial distributions of the internal magnetic field given by ${\bf B}_{int}={\hat z} \nabla_x A_y(x,y)$, in the homogeneous state (a) and the inhomogeneous state with vortices (b), where $B_{int}$ is normalized by $(\hbar c/2e) \xi(T)^{-2}$ and the length are given in units of $\xi(t)$. \label{fig:flux}}
\end{center}
\end{figure}

In the homogeneous helical state, only one Fourier component $c_n$ whose wave vector corresponds to $q_y(+\infty)$, $n=-4=n_b$, is non-vanishing, so that $|\Delta|$ is uniform along the $y$-axis. The inhomogeneity of the RSOC is reflected not only in the $x$-dependence of $|\Delta|$ through inhomogeneous paramagnetic depairing effects (see Eq.(\ref{eq:Hc2})) but also in a large spatial variation of $ A_y (x)$. The limiting difference in $A_y$ between the two sides is given by
\begin{equation}
A_y (+ \infty) - A_y (- \infty) = \frac{\Delta \tilde{K}}{K} H  \frac{c}{2e} = \frac{\Delta \tilde{K}}{\hbar K} H \frac{\Phi_0}{2 \pi},
\end{equation}
where $ \Phi_0 $ is the standard flux quantum. The local current ${\bf j}={\bf j}_s+{\bf j}_L+{\bf j}_{em}$ is shown in Fig.\ref{fig:gap-current} (a). Near the domain boundary at $x=0$, the current flows along the $y$-direction with a vanishing net current because of the Meissner screening effect. Consequently, as shown in Fig. \ref{fig:flux} (a), the induced internal magnetic field is uniform along the $y$-axis, concentrated  to the vicinity of the boundary on the length scale $ \lambda $ and oriented along the $z$-axis. The out-of-plane magnetic flux located around the boundary is given by $ \phi = A_y (+ \infty) - A_y (- \infty) $ per unit length of the boundary.

The net magnetic flux is reduced for the other state when spatial dependence of $ |\Delta|$ is introduced along the boundary ($y$-axis). This means that several Fourier components in Eq.(\ref{eq:boundary}) are non-zero. For our choice of parameters 
the gap function is a superposition mainly of the components $c_{-2}$ and $c_{-4}$ whose wave vectors $n=-2$ and $n=-4$ correspond to different helical vectors on the two sides $q_y(-\infty)$ and $q_y(+\infty)$, respectively (Fig.\ref{fig:x-dep}). This reflects directly different RSOC and, thus, different shifts of the Fermi surfaces in the two half-planes \cite{Springer}. 
In the extreme case we expect here $ A_y (+ \infty) - A_y (- \infty) = 0 $, such that  
\begin{equation}
\Delta q = q_y (+ \infty) - q_y (-\infty)= \frac{\Delta \tilde{K}}{\hbar K} H.
\end{equation}
Under this condition the two helical states have to be matched at the boundary which leads to a regular array of
topological defects, corresponding to vortices lined up along the boundary. 
The spatial structures of $|\Delta|^2$ and ${\bf j}$ are shown in Fig. \ref{fig:gap-current} (b). The vortices are visible through circular currents centered around the vortex cores where $|\Delta|$ is suppressed to zero. The internal field in this vortex state is shown in Fig. \ref{fig:flux} (b). This vortex pattern is superposed on the magnetic field of opposite sign spread along the boundary. Note that in Eq.(\ref{eq:b-source}) now both contributions to $ {\bf B}_{eff} \parallel \hat{z} $ are active. 

The appearance of vortices corresponds here to a screening effect. The vortices being opposite to the background magnetic field reduce the net magnetic flux on the boundary. Naturally, at low in-plane fields (''weak'' Fermi surface shift) the homogeneous helical state is stabilized and at a higher in-plane field a transition to the inhomogeneous state with vortices occurs. 

Insight on the relative stability of the two states can be gained through the free-energy density averaged along the $y$-axis shown in Fig. \ref{fig:x-dep} (c). Two prominent features can be identified, a rather localized contribution for the vortex phase (characterized through strong spatial variations of the order parameter at the boundary) and a more extended contribution due to the magnetic flux which is more extended in the homogeneous phase. Energetically the transition between the two phases is determined by the competition between the magnetic energy of the homogeneous phase and the vortex energy.

We provide here a simple estimate of the critical field for the transition between the two states. The energy of the homogeneous state is dominated by the magnetic field contribution approximately given by $ E_{(1)} \approx (\phi / 2 \lambda)^2 (\lambda / 4 \pi) $. The vortex state contributes mainly through the energy of each vortex $ \epsilon_v $. Per unit length along the boundary the energy corresponds to
$ E_{(2)} \approx \epsilon_v / \ell_v $, where $ \ell_v $ denotes the distance between the vortices and is derived through $ \Delta q = 2 \pi / \ell_v $, i.e. each vortex includes $ 2 \pi $ phase mismatch between the helical states of the two sides. 
Comparing these two energies yields the criterion
\begin{equation}
H   \Phi_0^2 \Delta \tilde{K}   =  32 \pi^2 \hbar K \epsilon_v \lambda .
\label{hc-lambda0}
\end{equation}
With the $H$-dependence of $\epsilon_v$ and $\lambda$ through $|\Delta|^2$,
we obtain the critical field for the discontinuous transition,
\begin{equation}
H_{c}(T) =  \frac{H_{p,-}(T)}{\sqrt{1+\beta_\lambda^2}}, \quad \beta_\lambda=  \frac{\pi \hbar \, \lambda_0}{{\bar \epsilon}_v} \Delta {\tilde K} \, H_{p,-}(T), 
\label{hc-lambda}
\end{equation}
where $\epsilon_v=|\Delta|^2 \, {\bar \epsilon}_v$ and 
$\lambda_0=\lambda(|\Delta(H=0)|)$. 
Note that the spacing between the vortices is roughly proportional to the in-plane field $ H $.  

The width $d$ of the boundary has not appeared in our estimate, as we have assumed $ d \ll \lambda $ in our
approximation of $ E_{(1)} $. If $ d $ becomes comparable or even larger than $ \lambda $, the magnetic field distribution is spread out on the length $d$ and $ E_{(1)} \sim  (\phi / 2 d)^2 (d / 4 \pi) $ such that for $ H_c $, $ \lambda $ has basically to be replace by $d$ in Eq.(\ref{hc-lambda0}). Thus, the critical field increases.

We address now some experimental aspects to detect the features discussed above. A direct way to observe the magnetic pattern introduced by the in-plane field is the local detection of out-of-plane magnetic fields (e.g. scanning SQUID microscope). Note that the sign of the magnetic flux depends on the gradient of RSOC modulation. An island of stronger RSOC has boundaries of opposite magnetic flux patterns. No magnetic flux is generated in the region where the gradient of the RSOC is perpendicular to the applied in-plane field. A further signature of modulated RSOC is the inhomogeneous nucleation of superconductivity for an in-plane magnetic field which could be observed through resistivity measurements as a broadening of the superconducting transition. Moreover, we would like to address the critical field for the vortex state generation. Looking at the $ H $-dependence of the free energy, we find that the in-plane (spin) magnetization would grow roughly linearly with $ H $ in the homogeneous case and would discontinuously drop to a basically constant value for the vortex phase. 
     
Finally, we would like to comment that as one can see in Eq.(\ref{eq:b-source}), not only inhomogeneous RSOC would generate the flux pattern we discussed, but also other inhomogeneities such as non-uniform in-plane fields would have a similar effect. 
Moreover, we would like to mention that twin-boundaries in non-centrosymmetric crystals could yields similar properties in
a magnetic field \cite{iniotakis}. 
In view of the fact that spontaneous ferromagnetism had been reported for
the LaAlO${}_3$-SrTiO${}_3$ interfaces, an in-plane magnetic field may not be necessary to generate  flux pattern \cite{imaging}. Generally,
the configuration discussed here represents a good way to detect features of the helical phase. 

We are grateful to D.F. Agterberg, J. Goryo, H. Ikeda, V.P. Mineev and T. Neupert for helpful discussions. This work is supported by the Swiss Nationalfonds, the NCCR MaNEP, and the Pauli Center for Theoretical Studies.


\begin{thebibliography}{50}
\bibitem{pairing} Both spin and orbital pairing symmetries do not affect the result qualitatively as far as only one channel is dominant in the pairing interaction.
\end{thebibliography}
\end{document}